\documentclass[10pt,a4paper]{article} 
\usepackage[dvips]{graphicx}
\usepackage{amsfonts,amsmath,amssymb}
\usepackage{hyperref}
\usepackage{cite}
\usepackage{enumitem}
\numberwithin{equation}{section}
\newcommand{\bse}{\begin{subequations}}
\newcommand{\ese}{\end{subequations}}

\begin{document}
\title{\bf Long-term Oscillations and Universal Behavior in Pulsed Electric Fields}

\date{}
\maketitle
\vspace*{-0.3cm}
\begin{center}
{\bf Leila Shahkarami$^{a,1}$, Farid Charmchi$^{a,2}$}\\
\vspace*{0.3cm}
{\it {${}^a$School of Physics, Damghan University, Damghan 41167-36716, Iran}} \\
\vspace*{0.3cm}
{\it  {${}^1$l.shahkarami@du.ac.ir, ${}^2$farid.charmchi@gmail.com}}
\end{center}

\begin{abstract}
We thoroughly analyze the response of the zero-temperature ${\cal N}=2$ super Yang-Mills theory to time-dependent electric field quenches via holography. 
We specially focus on transient pulse-like configurations for the electric field, characterized by some model parameters, such as the maximum value of the electric field $E_0$ and the ramping time $\delta_t$ which determines the time interval for switching the electric field on and off. We also compare some of the results with those of tanh-like quenches. The term tanh-like quench is used for a quench that rises from zero to a final finite value during a finite amount of time.
Our numerical solutions demonstrate that when the system is subjected to pulse-like electric field quenches, the emerged electric current as a response goes through three stages as time passes. After excitation and rapidly-damping oscillatory stages, it experiences a long-lasting periodic oscillatory region. In fact the effect of the electric pulse remains in the system much longer than the duration of the presence of the electric field itself. It is extremely interesting that, as confirmed by power spectrum diagrams, these oscillations have a unique obvious frequency which is independent of the details of the electric pulse function and its parameters. Moreover, we observe a universal behavior in the adiabatic limit, when the ramping time tends to infinity.
In this limit, the early and late time dynamics of the response electric current does not depend on the time dependence of the electric field. In particular, we see that for both pulse-like and tanh-like quenches, the maximum value at the first peak of the oscillations approaches the static value of the current induced by the presence of a static electric field $E_0$. 
However, the fast quench behavior differs extremely for different kinds of quench functions.
\end{abstract}
\section{Introduction}
The quantum vacuum, rather than being empty, can be encountered as a nonlinear medium with fluctuating particles. Therefore, studying the effect of various strong fields on the vacuum could be a key towards understanding fundamental physical processes. One of the most important phenomena in this regard called the Sauter-Schwinger (or Schwinger) effect refers to the creation of electron-positron pairs from the vacuum by a constant homogeneous external electric field. A remarkable fact is that instead of being restricted to QED, this effect is a universal feature of the quantum vacuum in the presence of a U(1) gauge field. Since quarks have electric charges, pairs of quark and antiquark can be produced if the applied electric field is strong enough to overcome the confining force sticking the quarks together and lead to a vacuum decay.

The calculation of the pair production rate from the QED vacuum under the influence of a constant electric field was first performed based on a field-theoretical approach at small-coupling and weak-field conditions \cite{Schwinger} and then generalized to the arbitrary-coupling case \cite{manton1}. In the latter work, they found that when the electric field strength exceeds a critical value, a vacuum cascade is observed; the charged particles are free to be produced without any obstacles. However, since the Schwinger limit is higher than the validity region of the weak-field condition, this approach loses its power to predict such a critical behavior.

It would be very instructive if we could direct our theoretical investigations, even one step closer to the real situations in experiments. To that purpose, it is more realistic to employ electric fields which vary in time and even space instead of being constant and uniform. In recent years, much work has been done dealing with the impact of time- and/or space-dependent electric fields on the quantum vacuum using different approaches. Among them we can mention applying the semiclassical approach of the worldline instantons to the study of constant and inhomogeneous electric fields but in restrictive conditions of weak-field and weak-coupling \cite{semi2}, indicating that the dynamic electric fields result in the reduction of the critical electric field which is good news in the experimental point of view. Furthermore, one can follow the studies dealing with the periodic time dependence in \cite{periodic}, boosting the creation rate by superposing a strong slowly-varying field with a weak but rapidly-changing field in \cite{assist}, and other generalizations to include spatial gradients in \cite{critical}. Some other sample works treating the electric fields as space- and/or time-dependent using methods such as quantum kinetic equations, semiclassical techniques, Monte Carlo simulations, and scattering calculations  can be found in \cite{scatter1,semi1,kinetic2,kinetic1,exact1}.

The nonperturbative nature of this phenomenon and the strong electric field involved in this problem, makes us resort to methods valid and powerful in such situations. AdS/CFT correspondence \cite{Maldacena1,Maldacena2,Maldacena3,Solana} has proved itself as a really powerful method for studying field theories at strong-coupling limit such as QCD at low energy, where the standard perturbative methods lose their power in analyzing strongly-coupled systems. That is the reason why much work has been done in the context of the so-called holographic Schwinger effect, namely studying the Schwinger effect using AdS/CFT correspondence or gauge/gravity duality (or in general holography). Most of these works consider the effect of constant electric fields (see for example \cite{semenoff,potential,Sch1,Sch2,Sch3,confin1,confin2,Sch4,Sch5,Sch6,Sch7,
confinrev,dehghani,us,us2,me,Sch8}).

Holography can moreover be utilized for investigating the time evolution of out-of-equilibrium systems, the situation at which even a powerful method like lattice calculation is not efficient due to the sign problem. The power of holography in analyzing the out-of-equilibrium evolution of strongly-coupled systems under strong electric fields with no restriction in the value of the applied field, motivated the authors of \cite{decay1} to initiate such a study in holography. They employed an $\mathrm{AdS}_5 \times S^5$ geometry with a flat flavor D7 brane in the probe limit in the gravity side, describing the ${\cal N}=2~SU(N_c)$ supersymmetric theory at large $N_c$ and strong coupling, and with massless quarks. Then, they studied the real time dynamics of this system when a homogeneous electric field is turned on and reaches its maximum finite value in a finite amount of time.  They obtained the induced time-dependent electric current and by which studied the relaxation of the system to its final stationary state. 
One can find few other works \cite{decay2,equen,magneticdecay,aliakbari1,aliakbari2,aliakbari3,aliakbari4} following this paper. Despite the high potential of holography in describing the different situations regarding this effect which is both theoretically and experimentally important, the papers regarding time- and space-dependent electric fields are very rare. To the best of our knowledge, among holographic Schwinger effect studies, there is no work dealing with spatially inhomogeneous electric fields, due to the complications of numerical calculations. Moreover, regarding time-dependent electric fields, there are still many other questions yet to be analyzed using holography.

In this paper, we consider the response of the simplest gauge theory with a holographic dual to external time-dependent but spatially homogeneous electric field quenches using AdS/CFT correspondence. The electric field configurations are chosen to be pulse-like, i.e., raise from zero up to a finite maximum value during a finite amount of time and eventually drop again to zero after a time interval. Due to the importance of performing theoretical calculations for given laboratory conditions and settings, a vast amount of work, utilizing different techniques, but mostly in lower dimensions because of numerical and calculational difficulties, has been done employing field profiles with the spatial and/or temporal dependences chosen to be similar to the ones in experiments. They mostly chose configurations with realistic pulse shapes with a structure which might be relevant for the planned laser facilities. However, to our knowledge, lack of holographic calculations for electric field quenches with such temporal dependences is felt (some aspects of such electric quenches have been studied in \cite{aliakbari3}).

We shall study the simplest holographic model, namely the ${\cal N}=2~SU(N_c)$ super Yang-Mills theory at large $N_c$ and strong coupling which is dual to the ten-dimensional $\mathrm{AdS}_5 \times S^5$ geometry with an embedded probe flavor D7 brane. 
Then, we employ two different pulse-like profiles for the external electric field, and by numerically solving the equation found from the DBI action of the D7 brane, obtain the dynamical electric current produced as a response to the electric field.
Our focus is mainly on the long-term evolution of the resulting electric current. We moreover compare some of our results with the ones in the presence of tanh-like quenches which have been more studied with respect to pulse-like quenches in the context of dynamical holographic Schwinger effect.
By tanh-like quenches we mean the electric fields which start from zero and increase to a final nonzero value.

The remainder of the paper is as follows. In section 2 we briefly introduce the holographic setup of our interest. By embedding the probe D7 brane in the $\mathrm{AdS}_5 \times S^5$ geometry, we construct the D3-D7 brane configuration and then find the DBI action of the D7 brane equipped with a gauge field. Then, we present the equation governing the dynamics of the gauge field. The two sections following that are devoted to the investigation of the dynamics of the electric current found as a response to the applied time-dependent electric fields with different parameters and functions. Finally, we present the summary and concluding remarks.


\section{${\cal N}=2$ supersymmetric $SU(N_c)$ QCD at strong coupling}\label{backgr geo}
In this work our aim is to investigate the effect of electric field quenches of particular shapes on the ${\cal N}=2$ $\mathrm{SU}(N_c)$ supersymmetric QCD at large $N_c$ and infinite 't Hooft coupling at zero temperature. The gravity dual of this theory is a D3-D7 brane configuration in the $\mathrm{AdS}_5 \times S^5$ background geometry.
\subsection{D3-D7 brane configuration}
We shall start with the following ten-dimensional $\mathrm{AdS}_5 \times S^5$ metric produced by the near horizon geometry of $N_c$ D3 branes:
\begin{align}\label{metric1}
 ds^2_{10}&=\frac{R^2}{z^2}\left(- dt^2+d\vec{x}^2+dz^2\right)+R^2d\Omega^2_5.
 \end{align}
Here, $R$ is the AdS and the 5-shere radius. $z$ denotes the radial coordinate of the bulk and $z=0$ corresponds to the AdS boundary.
Moreover, the metric of $S^5$ is chosen in the form of $d\Omega^2_5=d\varphi^2+\cos ^2\!\varphi \,d \Omega^2_3+\sin ^2\!\varphi \,d\psi^2$.

To introduce fundamental quarks to the field theory, we embed D7 branes into the bulk of the gravity side and for simplicity we work in the probe limit to neglect the backreaction of the D7 branes on the geometry.
The DBI action describing the dynamics of the probe D7 brane is written as
\begin{align}\label{dbi}
 S_{D7}=-\tau_7 \int d^8 \xi 
\sqrt{-\det \left[P(g)_{ab}+2 \pi \alpha'F_{ab}\right]},
 \end{align}
where $1/\left(2 \pi \alpha'\right)$ is the fundamental string tension, $\tau_7=1/[g_s \alpha'^4 (2\pi)^7 ]$ is the tension of the probe D7 brane and $\xi^a$ represent brane coordinates.
$P(g)_{ab}=\partial_a x^M \partial_b x^N G_{MN}$ and $F_{ab}=\partial_a A_b- \partial_b A_a$ are, respectively, the induced metric and the electromagnetic field strength on the brane.
By embedding the probe brane along the directions ($t$,$\vec{x}$,$z$,$\Omega_3$), its configuration can be specified by determining the remaining transverse coordinates $\varphi$ and $\psi$ as functions of the brane coordinates.
Considering the symmetry, one can set $\psi=0$. 
Furthermore, since in the present paper we restrict ourselves to the case of massless quarks, a flat D7 brane would be the solution of the system\footnote{In the case of finite mass quarks under the influence of electric fields, one needs to consider coupled complicated equations. A clever numerical technique to deal with such equations, invented in \cite{fd1}, can also be found, e.g., in \cite{aliakbari4,equen,fd2}}, i.e., we can simply choose $\varphi=0$.
According to the AdS/CFT dictionary, the presence of an external electric field in one of the spatial directions of the field theory, say $x_1$, corresponds to a nonzero gauge field on the brane. Therefore, we write the gauge potential $A_a$ as
\begin{align}\label{A}
2 \pi \alpha' R^{-2}A_a d\xi^a=a(t,z)dx_1.
 \end{align}
Substituting the induced metric and the above gauge field into Eq.\,\eqref{dbi}, the DBI action and the resulting equation of motion are obtained as
\begin{align}\label{DBIind}
S_{D7}=-\tau_7 V_3 \Omega_3\int dt dz \ \frac{\sqrt{w}}{z^5},
 \end{align}
and 
\begin{align}\label{eq1}
\partial_z\left(\frac{\partial_z a(t,z)}{z\sqrt{w}}\right)-\partial_t\left(\frac{\partial_t{a}(t,z)}{z\sqrt{w}}\right)=0,
 \end{align}
where $w=1+z^4 \left(\partial_z a\right)^2-z^4 \left(\partial_t{a}\right)^2$.
Now, the only nonzero component of the gauge field can be divided into two parts as $a(t,z)=-\int^t E(t') dt'+h(t,z)$, where $E(t)$ is the external electric field applied in the $x_1$ direction of the field theory. 
With the aid of the near boundary expansion of the gauge field, according to the AdS/CFT dictionary, one obtains $h(t,z=0)=0$, $\partial_z h(t,z)|_{z=0}=0$, and $\partial_z^2 h(t,z)|_{z=0}\propto j(t)$, where $j(t)$ is the electric current in the field theory.
We analyze the response of the system to the external electric field through the time evolution of the electric current $j(t)$. 
Hence, we have to solve Eq.\eqref{eq1} using the appropriate conditions stressed above at $z=0$ plus two initial conditions at initial time $t_0$ which are $a(t_0,z)=0$, $\partial_t a(t,z)|_{t_0}=0$.
\section{Oscillating electric current}

\subsection{The ``$\mathrm{{\bf cosine}}^2$ pulse"}
\begin{figure}[h]
\begin{center}
\includegraphics[width=7cm]{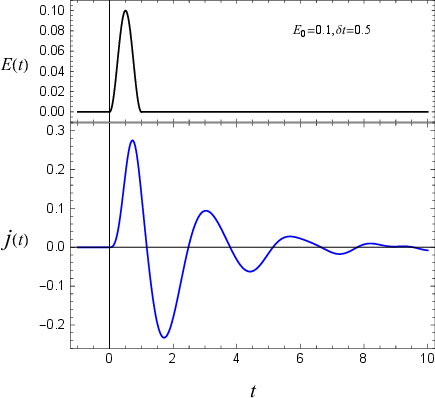}\hspace{1cm}
\includegraphics[width=7cm]{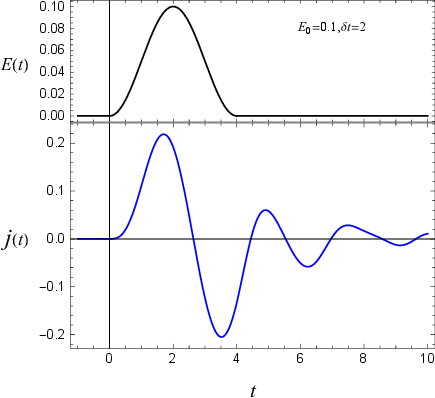}
\end{center}
\caption{\footnotesize 
Left (right) graph shows the response of the system, i.e., the electric current $j(t)$ (bottom) to a narrow (wide) $\mathrm{cosine}^2$ pulse (top).}
\label{pulse1andj}
\end{figure}
\begin{figure}[h]
\begin{center}
\includegraphics[width=7cm]{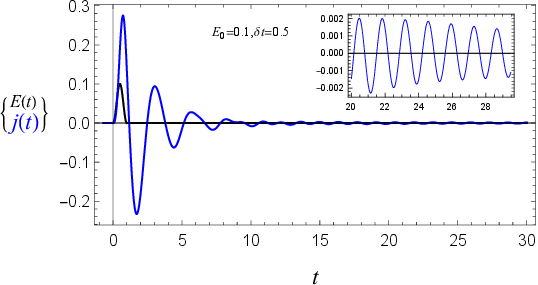}\hspace{1cm}
\includegraphics[width=7cm]{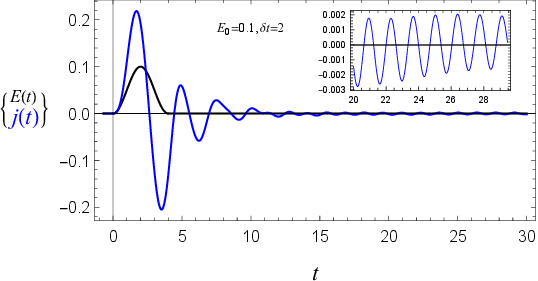}
\end{center}
\caption{\footnotesize 
The dynamical response of the system, $j(t)$, to the same $\mathrm{cosine}^2$ pulses as the ones in Fig.\,\ref{pulse1andj}, plotted over a larger time interval.}
\label{pulse1andjlong}
\end{figure}

 
We now proceed by specifying the time dependence of the applied external electric field. 
The hope for the observation of the Schwinger effect by high intensity lasers motivates the theoretical investigations to consider electric field configurations which are similar to laser pulses applied in real experiments and have the potential to realize the vacuum pair production.
To this purpose, we first investigate the effect of a time-dependent electric field in the form of
\begin{align}\label{pulse1}
E(t)=E_0 \begin{cases}
0,&t<0,\\
\cos^2\!\left(\frac{\pi t}{2\delta t}+\frac{\pi}{2}\right),&0\leqslant t\leqslant 2\delta t,\\
0,& t>2\delta t.
\end{cases}
 \end{align}
This function, which we call ``$\mathrm{cosine}^2$ pulse", has two free parameters: the ramping time $\delta t$ over which the pulse is turned on and off, and the maximum value of the electric field $E_0$.
Samples of this function for given parameters can be seen in the top graphs of Fig.\,\ref{pulse1andj}.
Moreover, the time evolutions of the electric current, as the dynamical response to the respective pulse functions, are depicted in the bottom graphs of the same figure. 
It can be seen that the first peak of the current in the left graph is closely, but with a small delay, following the focused narrow pulse. 
The delay in response can also be more clearly observed in the left graph of Fig.\,\ref{pulse1andjlong}, which plots the long-term evolution of the current $j(t)$ induced by the same pulse function. Here, the pulse is shown again by a black solid curve. The long-term evolution of the same current as in the right graph of Fig.\,\ref{pulse1andj} is also plotted in the right graph of Fig.\,\ref{pulse1andjlong}. From the right graphs of these two figures, one observes that when a wide pulse is applied, the system has enough time to respond and the first peak is completed before the completion of the pulse. Another observation is that the height of the first peak is lower when dealing with the wider pulse. We will elaborate further on this issue later.

In all cases, as can be obviously seen in Fig.\,\ref{pulse1andjlong}, the time-dependent electric current faces three stages. At first it experiences the excitation stage, when the system is driven out of equilibrium by the quench and the first peak begins. Indeed, the coupling of the matter field to the electric field makes the vacuum unstable and as a result an electric current emerges on the brane. Interestingly, after the first peak in $j(t)$, the electric current goes through an oscillatory stage (rapidly-damping oscillatory stage), with comparable amplitudes as the ones in the first peak. This stage lasts even after the electric field is reduced to zero again, at least for pulses of finite width.  
These oscillations are due to the polarization of the vacuum. In fact, the integral of the electric current over time is related to the vacuum polarization.
Notice that by increasing $\delta t$, the time interval of the mentioned stages is extended.
Here, our system will not eventually settle down into a thermalized state since no stable effective event horizon will form on the brane in contrast to the case in which the electric field is switched on from zero to a finite nonzero value \cite{decay1,aliakbari1,aliakbari2,aliakbari3,aliakbari4}, such as the following function:
\begin{align}\label{pulse3}
E(t)=E_0 \begin{cases}
0,&t<0,\\
\cos^2\!\left(\frac{\pi t}{2\delta t}+\frac{\pi}{2}\right),&0\leqslant t\leqslant \delta t,\\
1,& t>\delta t.
\end{cases}
 \end{align}
From now on, we call such quenches, tanh-like quenches. 
For the pulse we are studying here (given in Eq.\,\eqref{pulse1}), it can be observed that after the primary response, i.e.\,the first peak and the subsequent oscillations, is faded, the electric current enters a long-lasting oscillatory region. 
Similar behavior has also been noted in literature using nonholographic approaches (e.g., see \cite{kinetic2}\footnote{In this paper they use the framework of the quantum kinetic equations to consider the dynamical Schwinger effect in the context of $1+1$ QED. By incorporating the backreaction of the fermions (produced pairs) on the Maxwell field, they arrive at the conclusion that the vacuum looks like an eternally swinging medium in which the total electric field and pair number keep oscillating with a constant frequency and amplitude.}).
The insets in both panels of Fig.\,\ref{pulse1andjlong} show a slow damping trend for the amplitude of the oscillations. However, in all cases studied, the oscillations last as far as we can proceed with the numerical solution of Eq.\,\eqref{eq1}. 

\begin{figure}[h]
\begin{center}
\includegraphics[width=7cm]{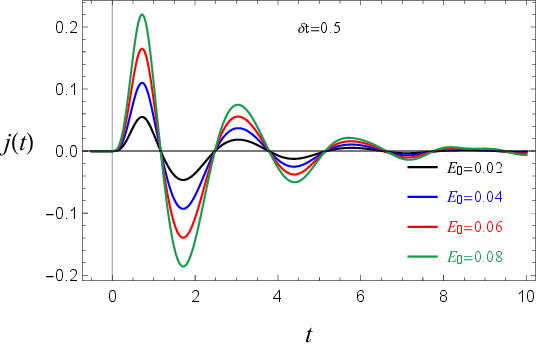}\hspace{1cm}
\includegraphics[width=7.3cm]{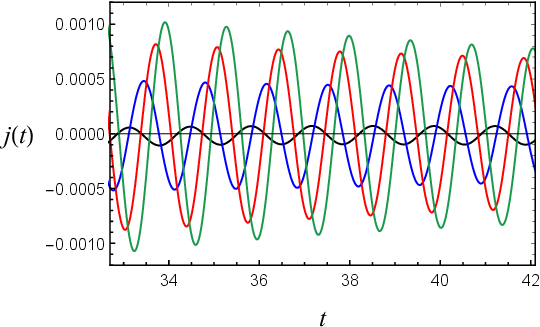}
\end{center}
\caption{\footnotesize 
Left (right) graph shows the early-time (late-time) dynamics of $j(t)$ for the $\mathrm{cosine}^2$ pulse for different values of $E_0$.}
\label{p1jtdvfix}
\end{figure} 

In Fig.\,\ref{p1jtdvfix} we study the effect of $E_0$ on the response of the system, when $\delta t$ is held fixed. Left (right) panel shows the early-time (late-time) response.
As can be seen, at fixed $\delta t$ the maximum value of the time-dependent electric field controls the amplitude of $j(t)$ in both early- and late-time regions.
At fixed $\delta t$, the time courses of the oscillations of $j(t)$, i.e., the times of the peaks and the times of zero-crossings, are exactly the same at early-time response as shown in the left panel. However,  this trend will not be held as the time passes (see the time courses in the right panel).

\begin{figure}[h]
\begin{center}
\includegraphics[width=7cm]{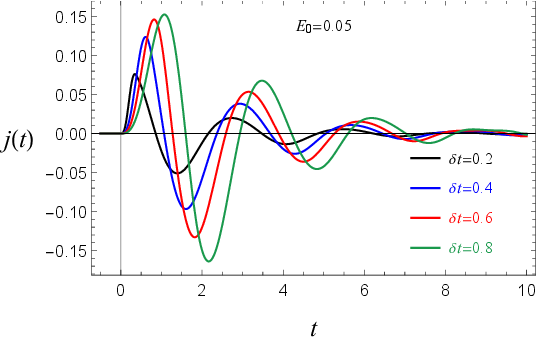}\hspace{1cm}
\includegraphics[width=7.3cm]{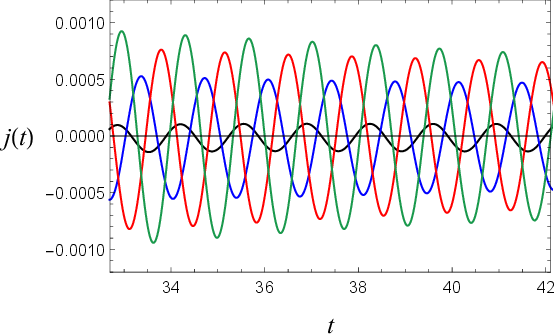}
\end{center}
\caption{\footnotesize 
Left (right) graph shows the early-time (late-time) dynamics of $j(t)$ when applying the $\mathrm{cosine}^2$ pulse, for different ramping times $\delta t$.}
\label{p1jteffix}
\end{figure} 

We continue by investigating the effect of varying the other parameter of the electric field quench, i.e., $\delta t$, at fixed $E_0$.
Figure \ref{p1jteffix} confirms that by increasing the ramping time $\delta t$, the time intervals of the first two stages are extended and hence the system enters the sustained oscillatory stage at a later time. 
Furthermore, an outstanding observation is that the increase of $\delta t$, increases the height of the oscillations in both early and late times. 
This is in contrast with the result extracted when the electric field is enhanced from zero to a finite value $E_0$ during the transition time $\delta t$, i.e., through a tanh-like quench such as the one given in Eq.\,\eqref{pulse3}. As argued in e.g., \cite{aliakbari1}, for such quenches the amplitude of the oscillations decreases by increasing $\delta t$. 
We will come back to this later.
 
\begin{figure}[h]
\begin{center}
\includegraphics[width=8.5cm]{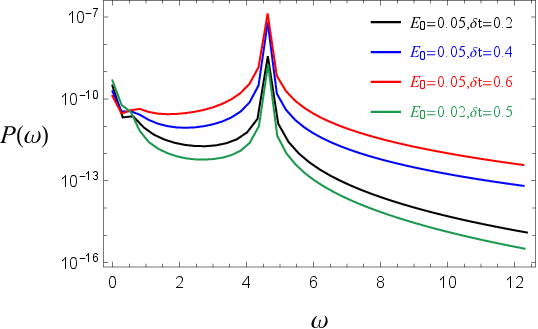}
\end{center}
\caption{\footnotesize 
The power spectrum $P(\omega)$ of $j(t)$ for sample $\mathrm{cosine}^2$ pulses. One can see a unique normal mode located at $\omega \approx 4.64$ corresponding to the periodicity ${\cal T} \approx 1.35$, in all cases.}
\label{p1ps}
\end{figure}

As mentioned, whenever the system is exposed to a pulse-like time-dependent electric field, i.e., a field that vanishes after a finite amount of time, the system eventually experiences a stage of long-term periodic oscillations.
Another highly interesting observation which could be extracted from the late-time behavior of the electric current displayed in the inset in Fig.\,\ref{pulse1andjlong} and the right graphs in Figs.\,\ref{p1jtdvfix} and \ref{p1jteffix} is about the frequency of the oscillations. As can be seen, the oscillating frequency seems to be the same, irrespective of the value of the parameters $E_0$ and $\delta t$.
This observation is confirmed by the power spectrum of the third stage of the electric current, $P(\omega)$, depicted in Fig.\,\ref{p1ps} for various values of the pulse parameters. Obviously, a single mode is recognized in all cases with the frequency $\omega \approx 4.64$ equivalent to the periodicity ${\cal T} \approx 1.35$. This universal frequency is the only normal mode of D7 brane oscillations, which remains in the late-time response of the system.

\subsection{The ``flat-top pulse"}
The Schwinger effect occurs in fact through a tunneling process in which the virtual charged particles gain required energy to become real particles. Therefore, for this effect to happen, the vacuum should be exposed to the electric field for a sufficient period of time. For this reason, we study the second pulse shape which we call the ``flat-top pulse". 
This pulse has two time scales, $\delta t$ and $\tau$. Over the ramping time $\delta t$ the electric field increases from zero to its maximum value, $E_0$. Then, for the finite amount of time $\tau$ the electric field remains constant at this value and falls off again to zero over the time $\delta t$.

\begin{figure}[h]
\begin{center}
\includegraphics[width=7cm]{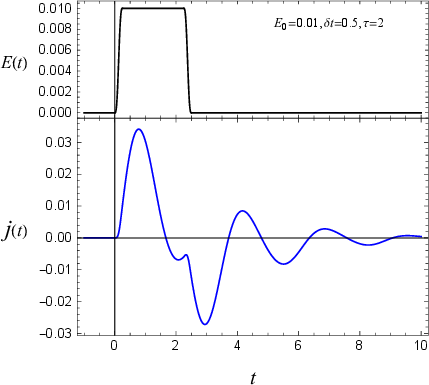}\hspace{1cm}
\includegraphics[width=7.2cm]{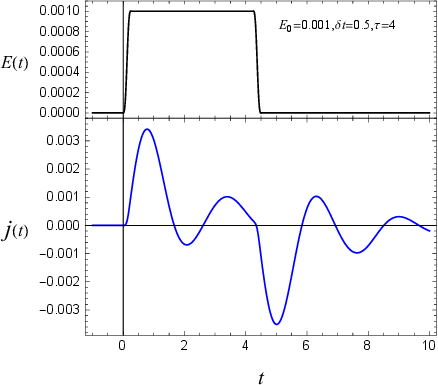}
\end{center}
\caption{\footnotesize 
The flat-top pulse (top) and the electric current $j(t)$ (bottom) versus time for various parameters of the pulse. 
}
\label{pulse2andj}
\end{figure} 

In Fig.\,\ref{pulse2andj} we exhibit the time dependence of the external electric field in the shape of flat-top pulse and the resulting electric current for various parameters of the pulse.
An interesting feature is that there appear similar electric currents but in opposite directions. When the electric field is turning on, a net flow of charged quarks starts in the same direction (positive direction) as the electric field. Due to the fluctuations forced by the evolving electric field, the electric current follows a damping oscillatory trend and if there were enough time $\tau$, it would eventually settle down in a thermalized state with the same current as the static case.  However, as soon as the electric field starts to fall down to zero again, the system is shocked again and the electric current follows a similar path but in the opposite direction (negative direction). These two opposite responses to the electric field while turning on and off, are more evident when the time duration $\tau$ is larger, as can be seen in the right panel of Fig.\,\ref{pulse2andj}.

\begin{figure}[h]
\begin{center}
\includegraphics[width=7cm]{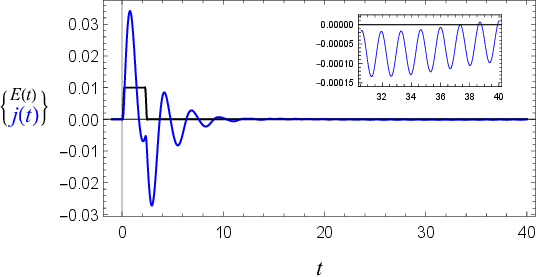}\hspace{1cm}
\includegraphics[width=7cm]{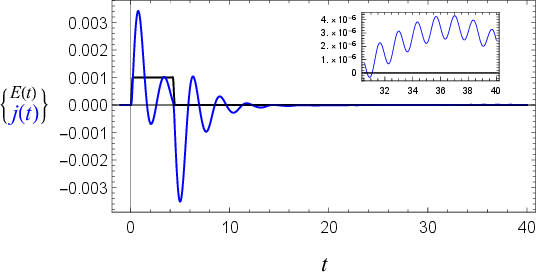}
\end{center}
\caption{\footnotesize 
The long-term response of the system, $j(t)$, to the same flat-top pulses as the ones in Fig.\,\ref{pulse2andj}.}
\label{pulse2andjlong}
\end{figure} 

\begin{figure}[h]
\begin{center}
\includegraphics[width=7cm]{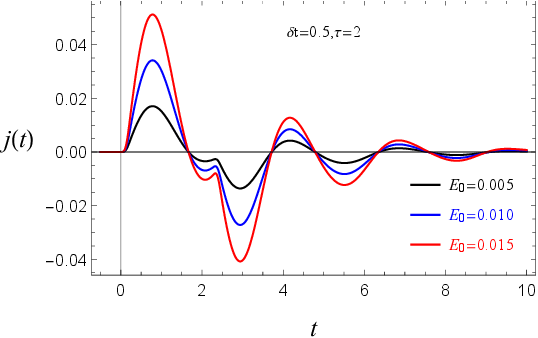}\hspace{1cm}
\includegraphics[width=7.3cm]{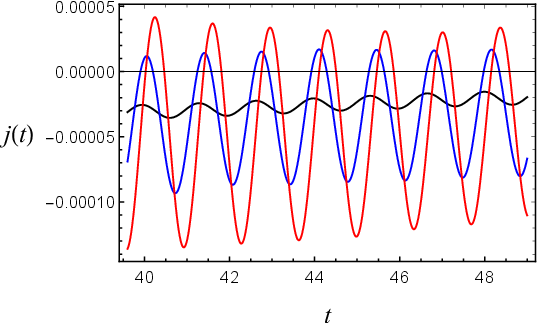}
\end{center}
\caption{\footnotesize 
Left (right) graph shows the early-time (late-time) dynamics of $j(t)$ for the flat-top pulse for different values of $E_0$.}
\label{p2jtdvfix}
\end{figure} 

\begin{figure}[h]
\begin{center}
\includegraphics[width=7cm]{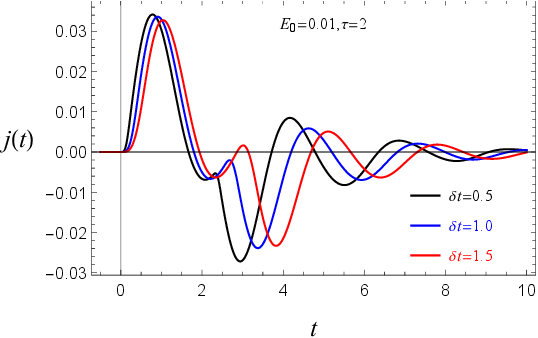}\hspace{1cm}
\includegraphics[width=7.3cm]{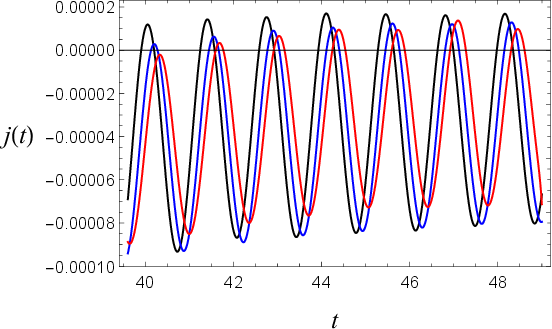}
\end{center}
\caption{\footnotesize 
Left (right) graph shows the early-time (late-time) dynamics of $j(t)$ for the flat-top pulse for different ramping times $\delta t$.}
\label{p2jteffix}
\end{figure}

Figure \ref{pulse2andjlong} exhibits the time evolution of the electric current extracted from the second derivative of the solution of Eq.\,\eqref{eq1}, over a long duration. The pulses represented by solid black curves have the same parameters as the ones in Fig.\,\ref{pulse2andj}. 
Interestingly, the response displays a similar long-term regime as in the previous case (the $\mathrm{cosine}^2$ pulse case).
After the electric pulse ceases, the electric current passes rapidly-damping oscillations and then enters the long-lasting oscillatory region.
These sustained oscillations can be better seen in the right graphs of Figs.\,\ref{p2jtdvfix} and \ref{p2jteffix}.
Figure \ref{p2jtdvfix} illustrates the electric currents emerged as responses to flat-top pulses with the same values of $\tau$ and $\delta t$, but different $E_0$.
As can be seen, increasing the maximum value of the electric field enhances the amplitude of the oscillations, and thus leads to an increase in the vacuum polarization, as expected.
Furthermore, the time courses of the oscillations of $j(t)$ for different values of $E_0$ are synchronized at early time. However, a shift in oscillations is seen at late-time dynamics. 
We also compare the dynamics of $j(t)$ when the system is exposed to flat-top pulses with different values of the ramping time $\delta t$. The result of this comparison can be seen in Fig.\,\ref{p2jteffix}, where we observe that the increase of $\delta t$ delays the response of the system and also reduces the height of the oscillations at both early and late time. 

A notably important point here is that again there exists a seemingly single frequency in the late-time dynamics of the electric current in response to the flat-top pulse. It is extremely interesting that the normal mode frequency obtained by virtue of the power spectrum is the same as the one found for the $\mathrm{cosine}^2$ case. In fact, we conclude that the frequency of the normal mode of the D7 brane fluctuations is independent of the details of the applied pulse, including its functionality and the values of the parameters $E_0$, $\delta t$, and $\tau$.

\section{Fast and slow quenches}\label{s3}
This section is devoted to a further consideration of the impact that the value of the ramping time has on the response of the system.
One might think that keeping the maximum applied electric field fixed, the sharper the quench or equivalently the smaller the ramping time, the higher the first peak of the response electric current. This is in fact the case when we quench an electric field which grows from zero up to a finite constant value and stays there forever (see, e.g., Eq.\,\eqref{pulse3}). Different quench functions of this type have been studied in holographic dynamical Schwinger effect. The main advantage of employing such quench configurations is that they provide enough time for the system to go through the out of equilibrium stage and eventually settle down into a nonequilibrium steady state with the electric current which coincides with the response of the static case with constant electric field $E_0$. As a result, one can study the process of thermalization of the system under the influence of an electric field quench and find quantities like the thermalization time and deconfinement time which are of particular interest in connection with experiments in RHIC and LHC, for example.

\begin{figure}[h]
\begin{center}
\includegraphics[width=7cm]{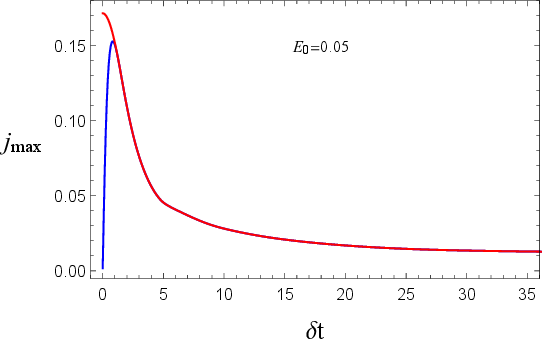}\hspace{1cm}
\includegraphics[width=7cm]{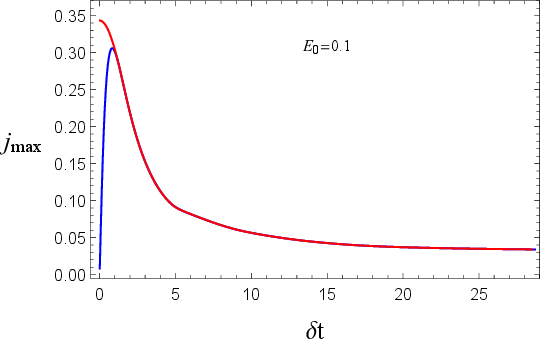}
\end{center}
\caption{\footnotesize 
The maximum value of the first peak of $j(t)$ versus $\delta t$, for the electric field function given in Eq.\,\eqref{pulse1} (blue curve) and Eq.\,\eqref{pulse3} (red curve), for two different values of $E_0$.  
}
\label{jmax}
\end{figure}

\begin{figure}[h]
\begin{center}
\includegraphics[width=7cm]{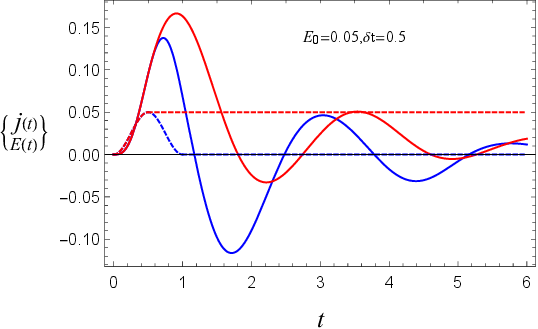}\hspace{1cm}
\includegraphics[width=7cm]{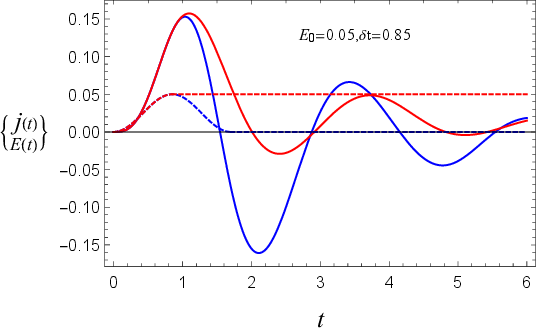}\\
\includegraphics[width=7cm]{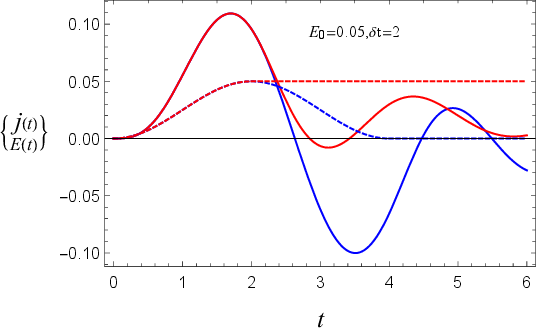}\hspace{1cm}
\includegraphics[width=7cm]{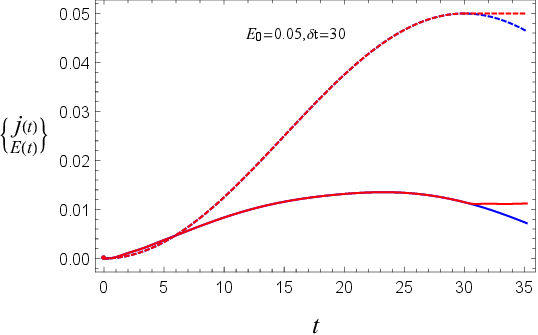}
\end{center}
\caption{\footnotesize 
The solid blue (red) curve shows the response to the electric field given in Eq.\,\eqref{pulse1} (Eq.\,\eqref{pulse3}). The time-dependent electric field in Eq.\,\eqref{pulse1} (Eq.\,\eqref{pulse3}) is also depicted by the dashed blue (red) curve. 
}
\label{jpulsetanh}
\end{figure}

As stated above, and also mentioned in literature (e.g., \cite{aliakbari1}), when the electric field with tanh-shape is quenched faster, i.e., the transition time $\delta t$ is smaller, the response of the system is more intense.
However, as can be realized from Figs.\,\ref{pulse1andj} and \ref{p1jteffix}, this is not in general the case when we are dealing with pulse-like quenches which are switched off after a finite amount of time. To learn more about this, in Fig.\,\ref{jmax} we illustrate the maximum value of the first peak of the time-dependent electric current as a function of $\delta t$. The blue (red) curve is related to the electric field given in Eq.\,\eqref{pulse1} (Eq.\,\eqref{pulse3}). As demonstrated in this figure, when we are dealing with an electric field similar to Eq.\,\eqref{pulse3}, $j_{\mathrm{max}}$ decreases by increasing $\delta t$, until apparently it asymptotes to a finite value. However, the blue curve shows a rise at very small values of $\delta t$ until $j_{\mathrm{max}}$ reaches a maximum value, and then it follows the same trajectory as the one in the red curve. It should be emphasized that the above explained behavior, i.e., the one in the blue (red) curve, is generic for every pulse-like (tanh-like) electric field, regardless of its exact functionality.
Contrary to the red curves, it seems that $j_\mathrm{max}$ as a response to $\mathrm{cosine}^2$ pulse (sketched by the blue curves) starts from zero and increases by increasing $\delta t$ until it reaches a maximum value 
and then falls off by further increasing $\delta t$. The smallness of $j_\mathrm{max}$ at small values of $\delta t$ seems to stem from 
the delay of the system in responding to the shock applied to the system. The pulse is switched off before the system has the chance to respond to it. This can be seen in Fig.\,\ref{jpulsetanh}, where both the pulse and the response are drawn for $\delta t=0.5,0.85,2,30$ which in Fig.\,\ref{jmax} are before the maximum, at the maximum, after it and in the asymptotic tail, respectively. In all graphs of this figure, $E_0=0.05$. Another interesting point is that the maximum of the $j_\mathrm{max}$ diagrams for $\mathrm{cosine}^2$ pulse (blue curves in Fig.\,\ref{jmax}) occurs at a unique $\delta t$ which is independent of $E_0$.

Let us now consider the adiabatic limit which corresponds to very slow quenches for which $\delta t$ is made arbitrarily large.
As can be seen in Fig.\,\ref{jmax}, regardless of the electric field functionality, the maximum value of the first peak approaches a finite value as $\delta t \to \infty$. Notice that this asymptotic value depends on $E_0$.
An interesting result is that this value coincides with the constant electric current obtained as the response of the system to a constant electric field $E_0$.
The asymptotic behavior is as if the system remains in a quasistatic equilibrium throughout applying the quench, in the adiabatic limit.

\section{Summary and conclusion}\label{conc}
In this paper, we have investigated the dynamical Schwinger effect, i.e., the creation of real charged particles from virtual ones induced by time-dependent electric fields.
The extension of the Schwinger effect studies to include the electric fields which are temporally and/or spatially inhomogeneous originates from practical and realistic purposes in the laboratory.
Therefore, recent studies mostly concentrate on the response of the system to the electric fields with the most similarity to focused pulses produced in currently operating laser systems.

Here, we have chosen two time-dependent pulse-like electric fields as quenches to ${\cal N}=2$ super Yang-Mills theory and studied its response, using AdS/CFT correspondence.
One of them called the $\mathrm{cosine}^2$ pulse is characterized by two model parameters; $E_0$ determining the highest value the electric field reaches and $\delta t$ which is the time required for increasing the electric field from zero to $E_0$ and vice versa. The other one called the flat-top pulse has an extra parameter $\tau$ determining the time interval during which the electric field remains fixed at $E_0$.
By numerically solving the equation of the gauge field introduced to the probe D7 brane embedded in the gravity side, we were able to extract the second derivative of the gauge field which is dual to the time-dependent electric current that flows in the gauge theory side as a response to the electric field.

By thoroughly examining the dynamics of the electric current $j(t)$ for various parameters, we have reached the results that we summarize here.
We have found that $j(t)$ experiences three stages: the excitation stage at which the electric current increases from zero in the same direction as the applied electric field. Then, the current enters an oscillatory evolution but with a rapidly damping amplitude. And finally, although the driving external electric field has been faded completely, the current $j(t)$  enters a sustained long-lasting oscillatory region. We should mention here that when we subject the system to a tanh-like electric field which is turned on from zero to a constant nonzero value during a finite amount of time, the electric current eventually relaxes to a nonequilibrium steady state. However, as illustrated in this work, by applying a short quench caused by a transient electric field pulse, the system does not return to its initial state at least for much more amount of time than the existence of the pulse itself. Instead, it experiences long-term oscillations. Another outstanding achievement is that these oscillations have a unique frequency as confirmed by the power spectrum of $j(t)$.
It is interesting that the frequency of the normal mode of the D7 brane fluctuations is independent of the functionality of the pulse and its parameters.
The observation of these oscillations is consistent with the oscillations noted in other research on Schwinger effect utilizing nonholographic methods.
Notice that even though concerning frequency, the response is unique regardless of the details of the applied pulse, the effect of the details of the pulse still remains in the late-time dynamics of the system as the amplitude of these long-term oscillations depends on the model parameters.

We furthermore have found that the increase of $E_0$, which is the maximum value of the pulse, enhances the amplitude of the oscillations and as a result the vacuum polarization, as is also the case with tanh-like quenches.
However, we have discovered a difference between the response to pulse-like and tanh-like quenches; although the amplitude of the oscillations decreases by the increase of the ramping time $\delta t$ of tanh-like pulses, the same is not always true in the case of pulse-like quenches.
By obtaining and drawing the maximum value of the first peak of $j(t)$, called $j_{\mathrm{max}}$, we have concluded that this value tends to zero for abrupt pulse-like quenches whereas for tanh-like quenches the greatest response occurs when $E(t)$ is turned on abruptly. By increasing $\delta t$, in the case of tanh-like quenches, $j_{\mathrm{max}}$ decreases until it reaches the static value of the response to a constant electric field $E_0$ at the adiabatic limit namely the infinitely slow quench with $\delta t \to \infty$.
However, in the case of pulse-like quenches, $j_{\mathrm{max}}$ increases by $\delta t$ at first till it reaches a maximum. By further increasing $\delta t$, it decreases similar to the tanh-like case and eventually approaches a constant value which is again the value of the resulting electric current as a response to the constant electric field $E_0$.
Another important point is that the maximum of $j_{\mathrm{max}}$ diagram occurs at a unique $\delta t$ regardless of the value of $E_0$.

As future directions, it would be interesting to study other holographic setups to clarify the relationship between the normal mode frequency of the brane fluctuations and different probably-related factors such as the induced geometry on the brane and the quark mass.
It also would be fruitful to extend our study to include massive quarks and/or holographic theories with a confining scale in the quark sector as future directions. These cases are involved with numerical difficulty and require much more efforts.




 \end{document}